\documentclass[superscriptaddress,showkeys,twocolumn,nofootinbib,longbibliography]{revtex4-1}

\usepackage{amsmath}
\usepackage{amssymb}
\usepackage{graphicx}
\usepackage{epsf}
\usepackage{slashed}
\usepackage{enumitem}
\usepackage[czech,english]{babel}
\usepackage[cp1250]{inputenc}
\usepackage{hyperref}
\usepackage{xcolor}
\usepackage{physics}

\usepackage[only,llbracket,rrbracket,llparenthesis,rrparenthesis]{stmaryrd} 
\usepackage{accsupp} 

\usepackage[font=small,labelfont=bf,justification=raggedright]{caption}
\usepackage{subcaption}

\usepackage{bbm} 

\usepackage[nodayofweek]{datetime}
\newdateformat{mydate}{\twodigit{\THEDAY}{ }\shortmonthname[\THEMONTH], \THEYEAR}
\usepackage[normalem]{ulem}

\usepackage{color}
\usepackage{ulem} 

\definecolor{myred}{rgb}{1,0,0}
\definecolor{mygreen}{rgb}{0,0.8,0.2}
\definecolor{myblue}{rgb}{0,0,1}

\definecolor{Ared}{rgb}{1,0.7,0}
\definecolor{Agreen}{rgb}{0.7,0.8,0.2}
\definecolor{Ablue}{rgb}{0,0.7,1}

\renewcommand{\emph}[1]{\textit{#1}}

\begin{document}


\title{Oscillons in the broken vacuum  and global vortex annihilation }

\author{D. Canillas Mart\'inez}
\affiliation{IUFFyM, University of Salamanca, Plaza de la Merced 1, 37008 Salamanca, Spain}

\author{A. Gonz\'{a}lez-Parra}

\affiliation{IUFFyM, University of Salamanca, Plaza de la Merced 1, 37008 Salamanca, Spain}

\author{D. Migu\'elez-Caballero}
\affiliation{Departamento de Fisica Teorica, Atomica y Optica and Laboratory for Disruptive Interdisciplinary Science (LaDIS), Universidad de Valladolid, 47011 Valladolid, Spain
}

\author{A. Wereszczynski}

\affiliation{Institute of Theoretical Physics, Jagiellonian University,
Lojasiewicza 11, Krak\'{o}w, Poland}
\affiliation{International Institute for Sustainability with Knotted Chiral Meta Matter (WPI-SKCM$^{\; 2}$), Hiroshima University, 1-3-1 Kagamiyama, Higashi-Hiroshima, Hiroshima 739-8531, JAPAN}

\begin{abstract}
In contrast to the complex $\phi^4$ model, vortex-antivortex collisions in the complex $\phi^6$ theory reveal a resonant structure due to the existence of a remarkably stable, long-lived, large amplitude oscillon in the broken vacuum. Surprisingly, it persists despite the absence of a mass gap associated with the flat direction in the broken vacuum. We demonstrate that its existence is related to a far-distance modification of the potential, namely, the appearance of an unbroken (false or true) vacuum. 
\end{abstract}

\maketitle

\section{Motivation}
Global strings \cite{Hindmarsh:1994re, Vilenkin:2000jqa} are hypothetical cosmic solitons that are conjectured to play a distinguished role in the evolution of the early Universe. E.g., they naturally appear in well motivated extensions of the Standard Model, where an additional Peccei-Quinn $U(1)$ symmetry is spontaneously broken. 
Of course, a detailed understanding of the dynamics of global strings is crucial for any application \cite{Shellard:1987bv, Yamaguchi:1998gx,Kawasaki:2018bzv, Hindmarsh:2019csc, Gorghetto:2020qws, Hindmarsh:2021vih, Blanco-Pillado:2022axf}. This especially concerns the realistic estimation of the relic abundance of axions, i.e., one of the most promising dark matter candidates \cite{Preskill:1982cy, Abbott:1982af}. 

A remarkable process in the evolution of the ensemble of cosmic
strings is the string-(anti)string collision, which, due to possible annihilation,  reduces the final number of the strings, releasing their energy into radiation of dark matter (and other) particles.
Indeed, in models with the $\phi^4$ potential, strings quickly annihilate,
transferring all of the energy to axionic radiation \cite{Shellard:1987bv}. 

Such a direct annihilation of the axionic strings is closely related to the non-existence of any medium- or long-lived intermediate states as, e.g., oscillons, in models with the complex $\phi^4$ potential. Oscillons are nontopological, ultimately unstable soliton-like excitations, which may have an extremely long life-time \cite{Bogolyubsky:1976nx, Gleiser:1993pt}; see also \cite{Amin:2010jq, Amin:2011hj, Zhang:2020bec, Olle:2020qqy,vanDissel:2023zva,  Li:2025ioq, Zhou:2024mea}. Although still mysterious, their origin and properties can be related to $Q$-balls \cite{Blaschke:2024dlt}, another class of nontopological solitons that are stabilized by the conserved $U(1)$ charge \cite{Friedberg:1976me, Coleman:1985ki}. 

The non-existence of oscillons in the complex $\phi^4$ theory is not surprising. In fact, it agrees with the common expectation that there are no oscillons in gapless theories, that is, theories where there is no mass threshold in the small perturbation problem. The Mexican-hat potential possesses a flat direction, which leads to the appearance of massless excitations, i.e., Goldstone modes. Any oscillon, if generically perturbed, should immediately decay via the emission of massless radiation \footnote{Oscillons in gapless models are known, but they exist in very special fine-tuned theories; see \cite{Dorey:2023sjh} and \cite{vanDissel:2025xqn}.}. 

The purpose of the current work is to investigate the impact of the inclusion of higher order self-interaction terms on the outcome of the annihilation process, and particularly on the appearance of long-lived oscillons which may change the overall budget of the produced radiation in the axion sector (we neglect fermions). Importantly, the generalizations considered here do not change the nature of the broken vacuum. It still has an $\mathbb{S}^1$ topology with a flat direction and, as a consequence, a vanishing mass threshold. What changes is the potential at a rather remote point $\phi=0$, where a local minimum or a new unbroken vacuum can develop. 

We will study this problem using a lower-dimensional version of a decoupled axionic string, {\color{red}} that is, global vortices \cite{manton2004topological} in a theory with the $\phi^6$ potential. Although this potential is not renormalizable, it is a low energy limit of a renormalizable theory with an additional real-scalar degree of freedom \cite{Friedberg:1976me, Dvali:1996xe, Dvali:2002fi} and, therefore, still relevant in various cosmological set-ups, see, e.g., its application in the erasure of vortices and monopoles \cite{Dvali:2022rgx, Bachmaier:2023zmq}. 

We will show that, while the scattering of global vortices in the $\phi^4$ theory is rather trivial, possessing only a direct annihilation channel, the $\phi^6$ theory surprisingly exhibits a very involved pattern of the vortex-antivortex (VAV) collisions. This complexity is attributed to the unforeseen emergence of long-lived \textit{oscillons} within the \textit{broken vacuum}.




\section{Global vortex models}
We will consider the global vortices in (2+1) dimensions defined by the following Lagrangian density 
\begin{equation}
\mathcal{L}=
\frac{1}{2} \partial_\mu \phi \,\overline{\partial^\mu \phi} -V\left(|\phi|\right),
\label{global}
\end{equation}
where we assume a family of potentials \cite{Christ:1975wt}
\begin{equation}
    V(|\phi|)=\frac{1}{8(1+\nu^2)} \left(\nu^2 + |\phi|^2 \right)\left( |\phi|^2 -2\right)^2,
    \label{eq:V_6}
\end{equation}
parametrized by $\nu \in [0,\infty)$.  The theory is invariant under the $U(1)$ global symmetry, $\phi \to e^{i\alpha} \phi, \, \alpha \in \mathbb{R}$. For all values of $\nu$, the potential has a broken vacuum $|\phi|=\sqrt{2}$, with $\mathbb{S}^1$ topology. 
This provides the global vortices, which are solutions of a static field equation arising from the ansatz $\phi=f(r)\,e^{in\theta}$
\begin{equation}
    \frac{d^2 f}{dr^2} +\frac{1}{r} \frac{df}{dr} -\frac{n^2}{r^2}f-V'_f(f) =0
\end{equation}
and supplemented by topologically nontrivial boundary conditions
$f(r=0)=0, \;\;\;f(r\to\infty)=\sqrt{2}$.
Here, we use the usual polar coordinates $r, \theta$. $n \in \mathbb{Z}$ is the winding number (vorticity). 

\begin{figure}
    \centering
   \hspace*{-0.5cm} \includegraphics[width=0.9\linewidth]{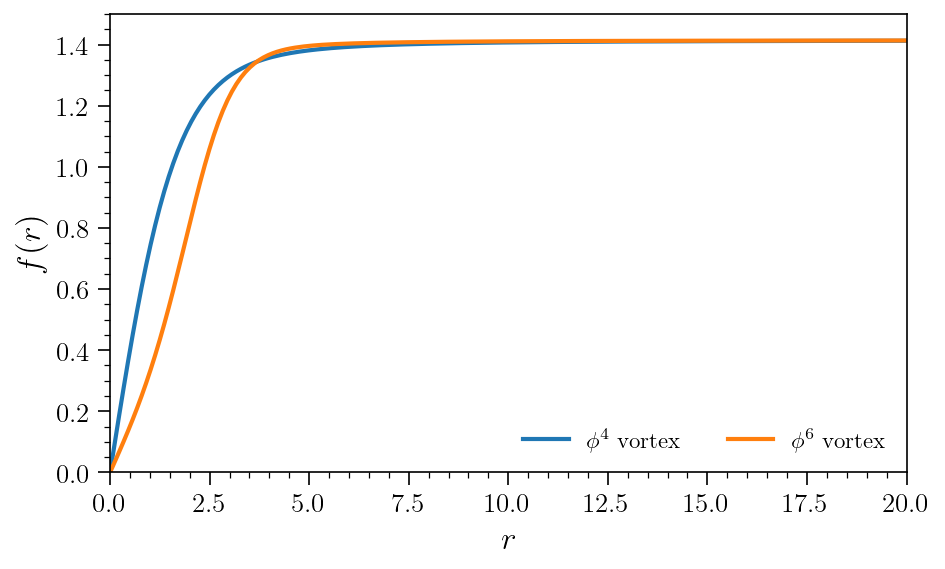}
    \caption{Profiles of the unit charge vortices in the $\phi^4$ (blue curve) and $\phi^6$ models (orange curve).}
    \label{fig:profiles}
\end{figure}

We will be mainly interested in the $\phi^6$ potential ($\nu=0$), which has an additional point-like unbroken vacuum at $\phi=0$. We will compare the results with the $\phi^4$ potential ($\nu \to \infty$), which  has a local maximum at $\phi=0$. 
In Fig. \ref{fig:profiles} we show the profile $f$ of the unit charge vortices. The vortices have infinite energy due to the large distance logarithmic divergence, which cancels for VAV solutions.   

As it was pointed out before, the broken vacuum has a flat direction which leads to massless excitations along the phase. This amounts to the existence of massless Goldstone's waves. In consequence, although the radial excitations are still massive with $m_{\text{br}}^2=4$, there is no mass gap in the two-channel perturbation problem in the broken vacuum. Hence, no oscillons are expected. 

Obviously, the unbroken vacuum of the $\phi^6$ theory does not have a flat direction. Thus, there is a mass threshold at $m^2_{\rm un}=1$ and, also, oscillons. 

\section{Vortex-antivortex collisions}
\begin{figure}
    \centering
   \hspace*{-0.5cm} \includegraphics[width=1.0\linewidth]{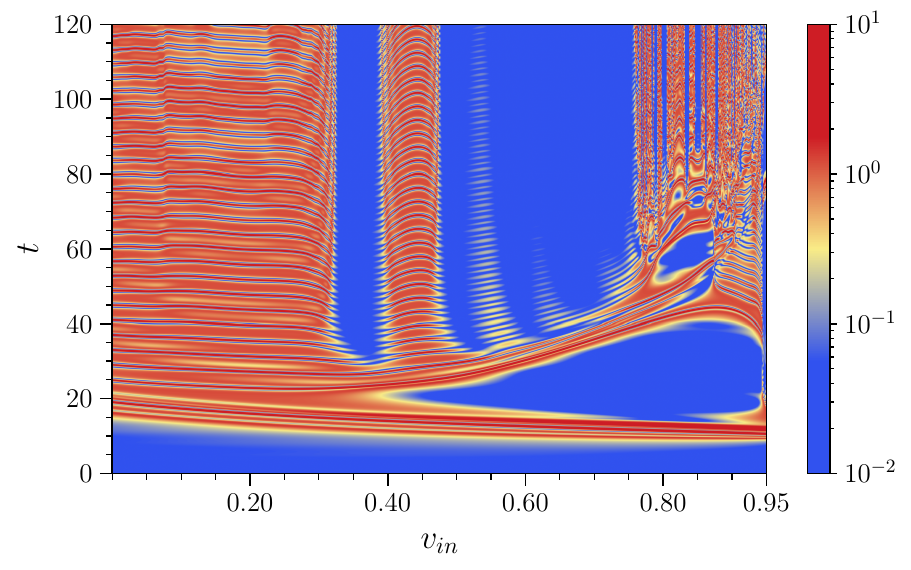}
   \hspace*{-0.5cm} \includegraphics[width=1.0\linewidth]{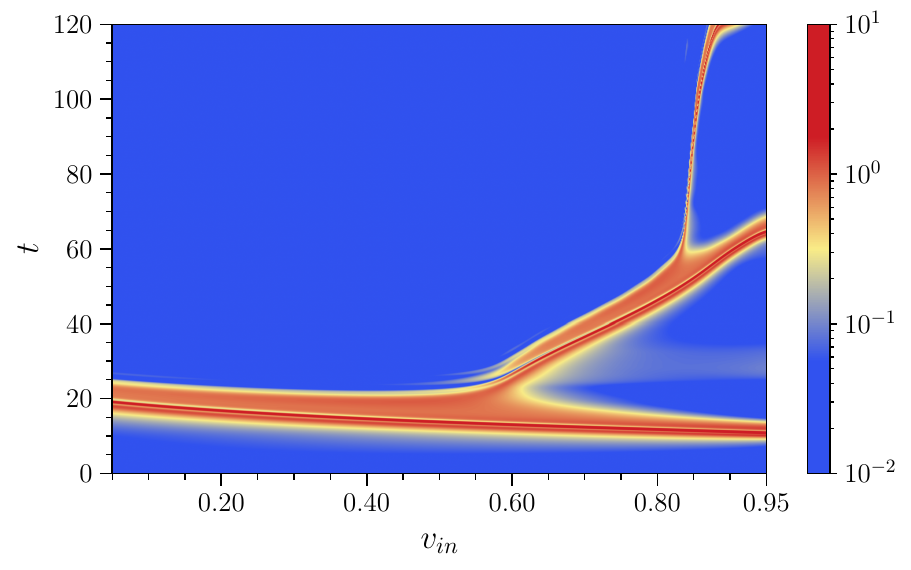}
    \caption{The VAV collision. Energy density at the origin as a function of time for various values of the initial velocity $v_{\rm in}$. Upper panel:  $\phi^6$ model. Lower panel: $\phi^4$ model.}
    \label{fig:scan}
\end{figure}

Now, we investigate the head-on collision of the VAV pair. In the initial state, the solitons are separated by $2d=20$, and boosted towards each other with initial velocity $v_{\rm in} \in [0,0.95]$. For numerical details, see Appendix \ref{numerical-setup}. Note that the vortices have long range tails and, therefore, the initial state, even for relatively large separation, describes a rather strongly perturbed VAV configuration, where the vortices significantly attract each other. As a result, particularities of the scattering depend on the initial separation, see Appendix \ref{VAV-d}. 

\begin{figure*}
    \centering
   \hspace*{-0.5cm} \includegraphics[width=0.25\linewidth]{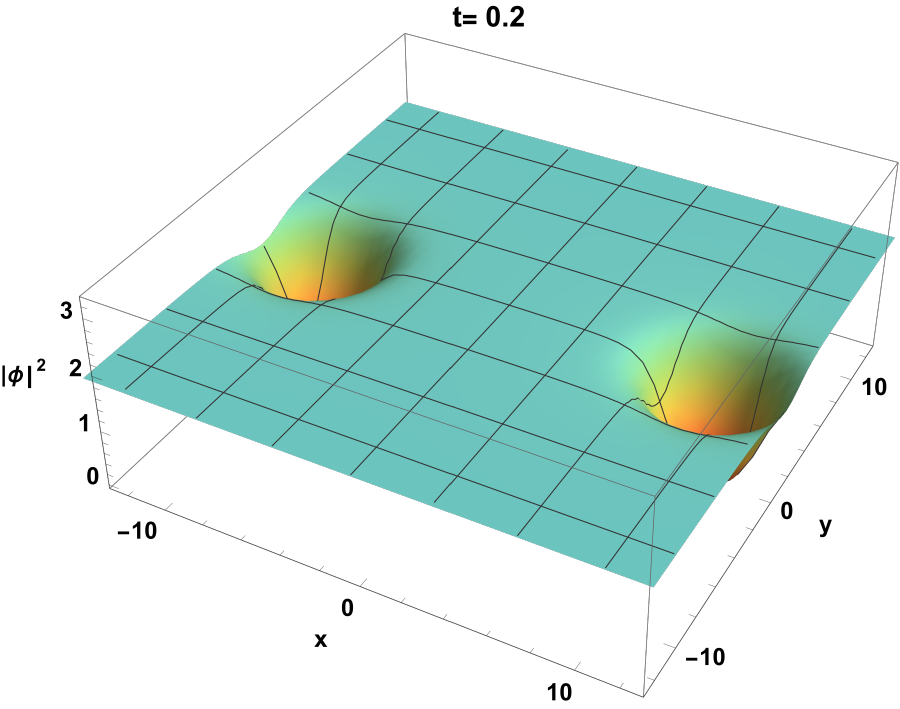}
   \includegraphics[width=0.25\linewidth]{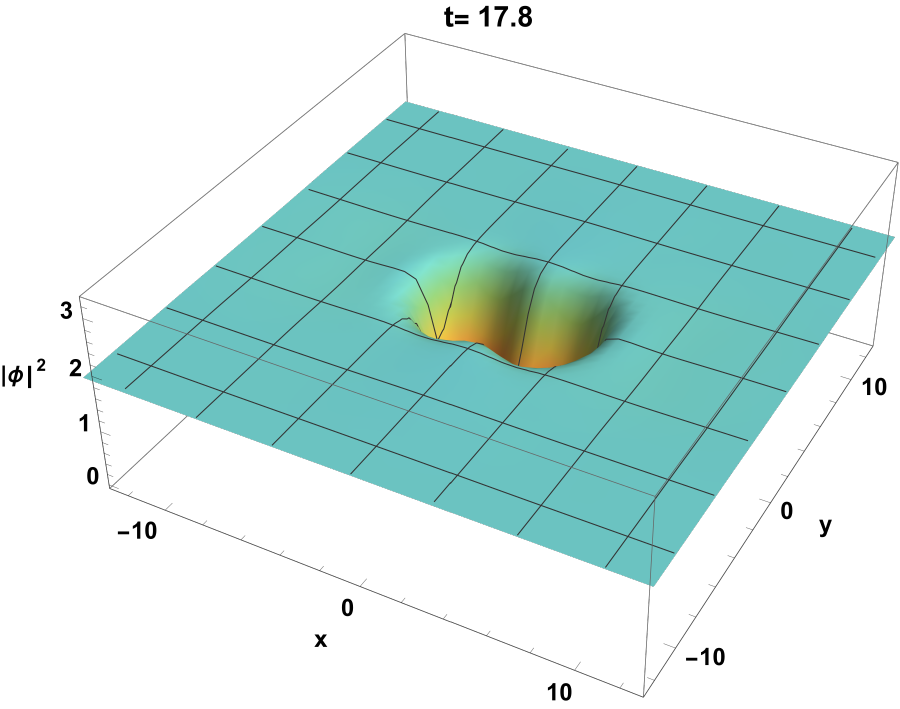}
   \includegraphics[width=0.25\linewidth]{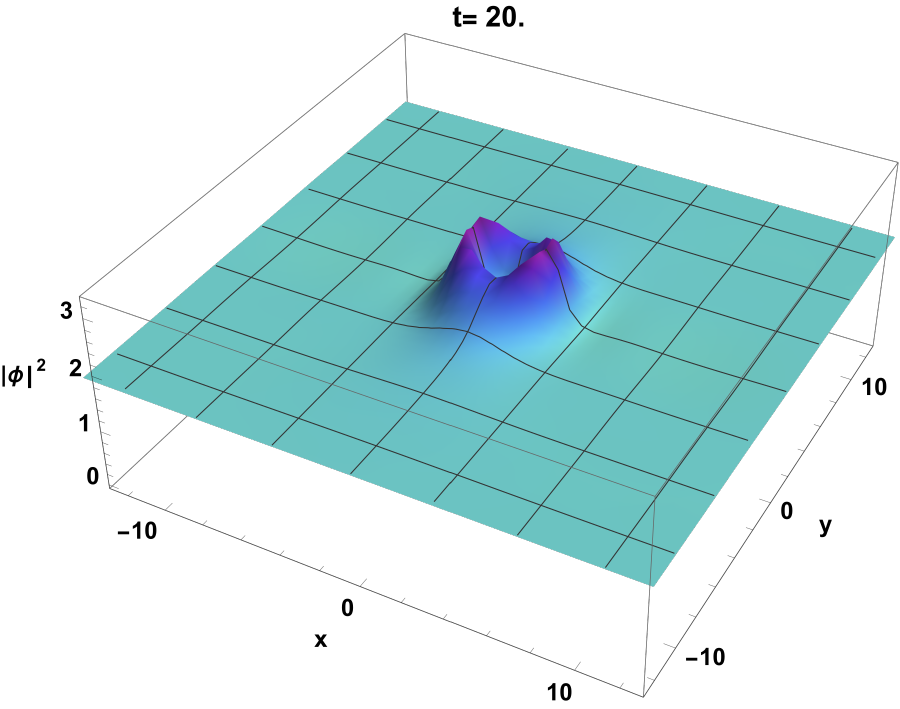}
   \includegraphics[width=0.25\linewidth]{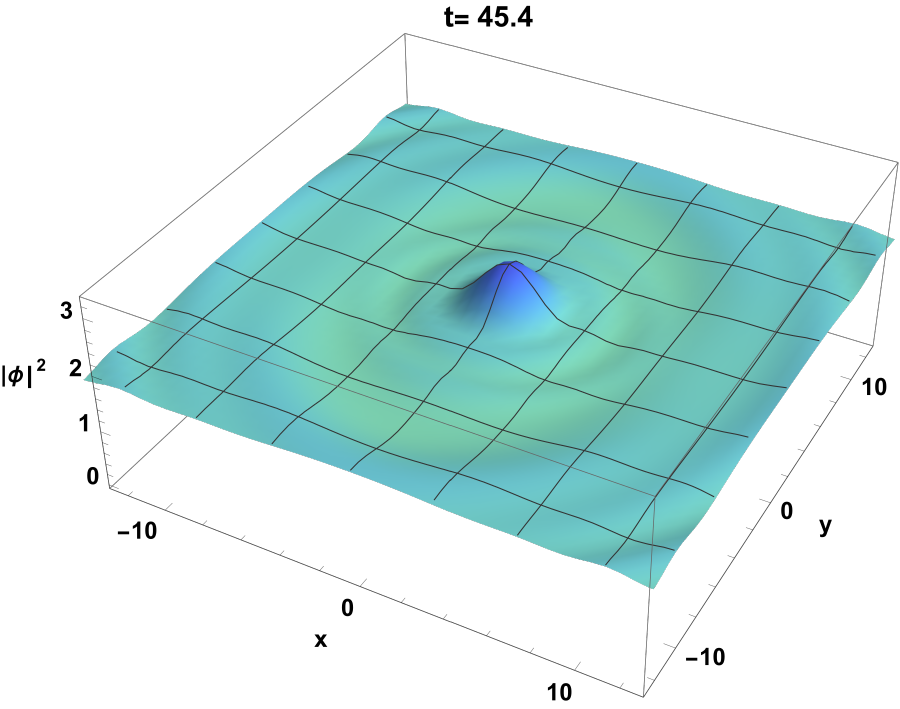}
   \centering
   \hspace*{-0.5cm} \includegraphics[width=0.25\linewidth]{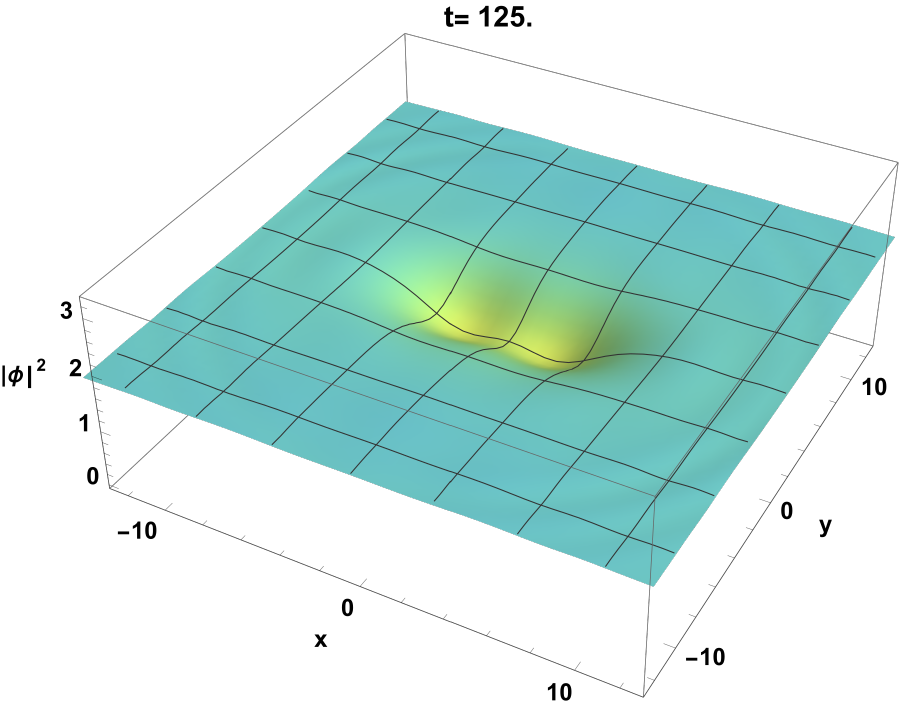}
   \includegraphics[width=0.25\linewidth]{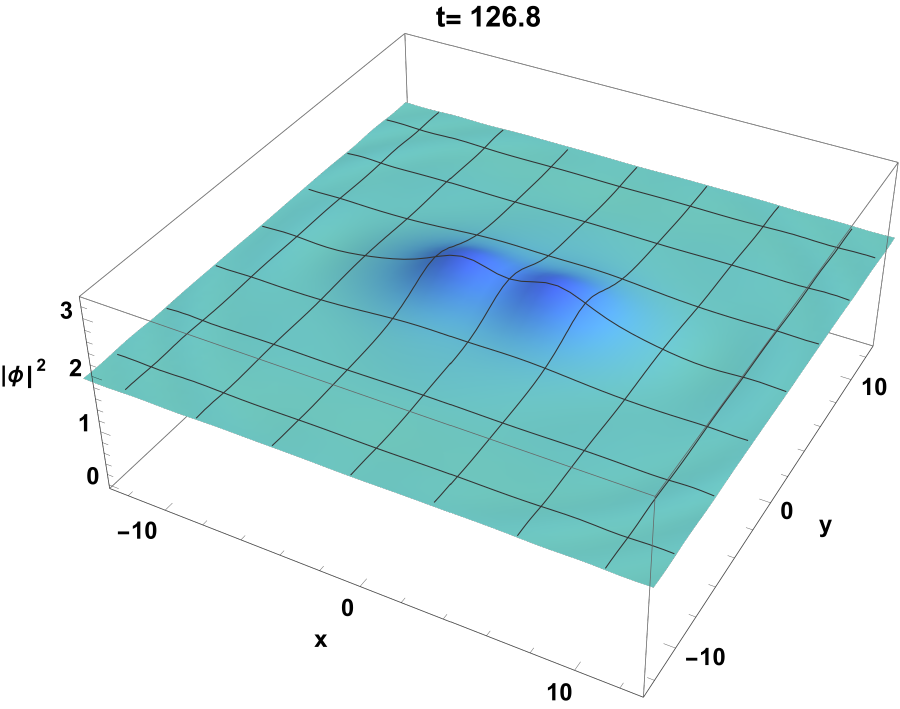}
   \includegraphics[width=0.25\linewidth]{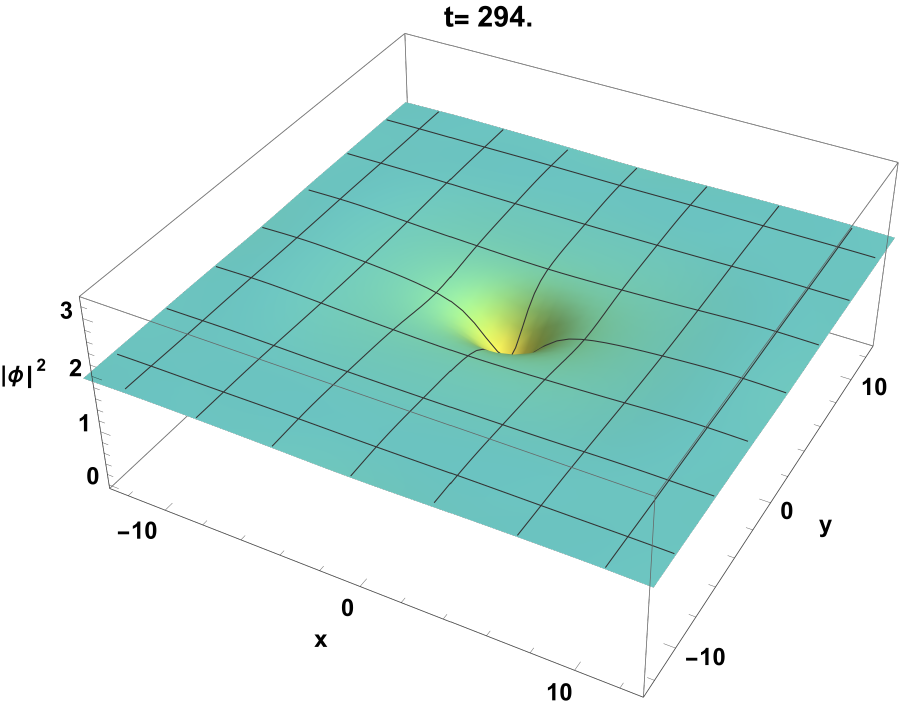}
   \includegraphics[width=0.25\linewidth]{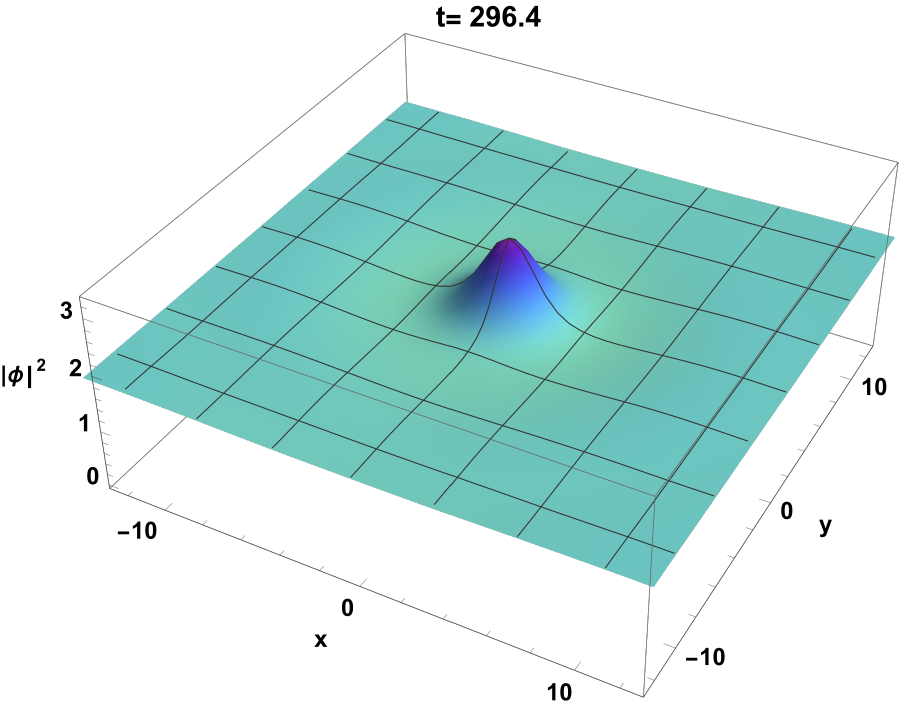}
    \caption{Formation of a radially symmetric oscillon in the VAV collision in the $\phi^6$ model with $v_{\rm in}=0$ and $2d=20$.}
    \label{fig:VAV}
\end{figure*}

The results for the $\phi^6$ model are shown in Fig. \ref{fig:scan}, upper panel, where we present the time evolution of the energy density at the origin as a function of $v_{\rm in}$. We see the characteristic chaotic resonant pattern in the final (actually, long-lived intermediate) state formation where quick annihilation scenarios occur with two, three, or more bounces. After each collision, the vortices pass through each other, Appendix \ref{VAV-bounce} (blue regions with a few red lines). We observe the formation of a long-lived, radially symmetric, oscillating structure, which corresponds to the oscillon stage (red regions). See Fig. \ref{fig:VAV}, where evolution with $v_{\rm in}=0$ is shown. 
Of course, the energy divergence implies that an infinitely separated vortex and  antivortex cannot exist in the final state

Such a resonant structure is very well known in the collisions of kinks \cite{Sugiyama:1979mi, Campbell:1983xu, Manton:2021ipk, Goodman, Takyi:2016tnc, Christov:2018ecz, AlonsoIzquierdo:2020hyp,Dorey:2017dsn, Blaschke:2023mxj,Bazeia:2022yyv, Mukhopadhyay:2023zmc, Marjaneh:2023dhu, Hahne:2024qby, Santos:2024ijd,Karpisek:2024zdj, Navarro-Obregon:2025xmw, Rafaj:2026uzz}, oscillons \cite{Blaschke:2024uec}, $Q$-balls \cite{Martinez:2025ana} and also local vortices \cite{Bachmaier:2025igf, Krusch:2024vuy, AlonsoIzquierdo:2024nbn, Alonso-Izquierdo:2025suz,AlonsoIzquierdo:2026mub}. It arises from the resonant energy transfer mechanism generated by internal modes hosted by scattered solitons \cite{Sugiyama:1979mi, Campbell:1983xu, Manton:2021ipk} or by modes of ephemeral excitations that appear as an intermediate stage of evolution \cite{Adam:2021fet, Martinez:2025ana}. In the present work, this structure is not a consequence of the global vortex modes (infinitely many such modes, see Appendix \ref{modes}), but is instead due to the oscillons. 

It is instructive to compare this with VAV scatterings in the $\phi^4$ model, where no resonant behavior or oscillon is found. We only observe  direct annihilation (after a few bounces for very high $v_{\rm in}$), see Fig. \ref{fig:scan}, lower panel. 
\begin{figure}
    \centering
   \hspace*{-0.5cm} \includegraphics[width=0.84\linewidth]{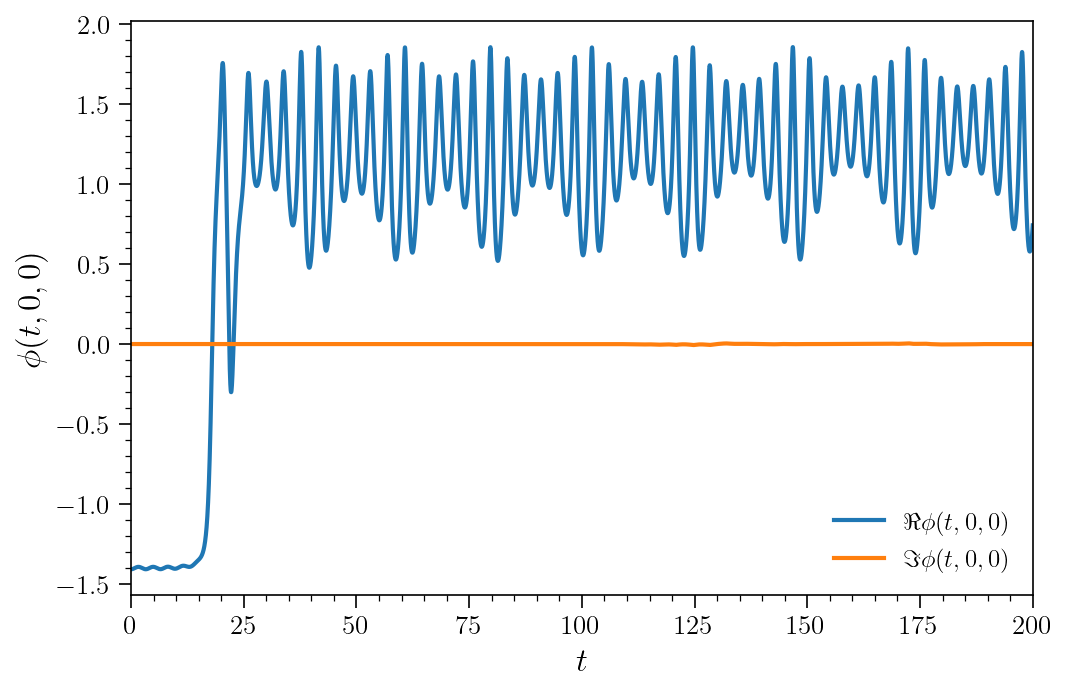}
    \centering
   \hspace*{-0.5cm} \includegraphics[width=0.84\linewidth]{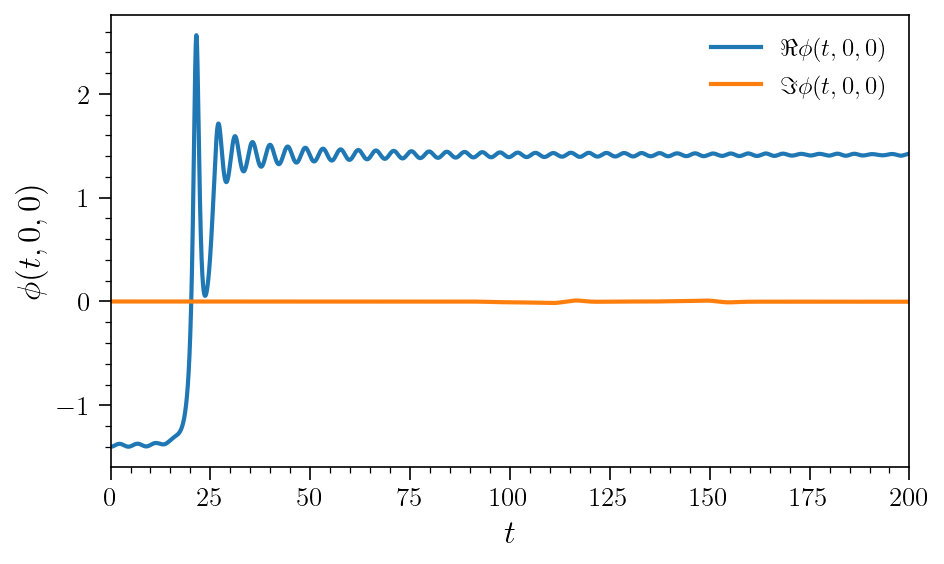}
    \caption{The real and imaginary component of $\phi$ at the origin in the VAV collision with $v_{\rm in}=0$ and $d=20$. Upper: oscillon formation in the $\phi^6$ theory. Lower: Absence of oscillon in the complex $\phi^4$ theory.}
    \label{fig:VAV-field}
\end{figure} 

To prove that an oscillon is formed, we analyze the field at the origin. We clearly see that, after the initially dramatic evolution, it quickly settles down to a remarkable stable state where the field has only the real radially symmetric component that performs radial breathing-like oscillations; see Fig. \ref{fig:VAV-field}, where we plot the real and imaginary component of the field at the origin for $v_{\rm in}=0$. In fact, the real component does not oscillate symmetrically around the broken vacuum. Additionally, we also checked that the energy stored in the imaginary part is quickly radiated, and this component of the field is basically zero everywhere. 

The frequency of the radial oscillations in Fig. \ref{fig:VAV-field} (upper panel) is remarkably stable, $\omega_0 = 1.863$, which is, as expected, below the mass threshold in the radial channel of small excitations, $m_{\rm br}=2$. It is also clearly visible that the oscillon has some amplitude modulations, which arise from the second (doubled) peak in the power spectrum located at $\omega_0\pm \omega_{\rm mod}$, $\omega_{\rm mod} = 0.132$. There are also higher harmonics. As proposed in \cite{Blaschke:2024dlt}, such a modulated large amplitude oscillon is a bound state of two unmodulated oscillons.

However, it should be noted once again that this oscillon lies in a broken vacuum that lacks a mass threshold due to the existence of a massless Goldstone excitation along the flat direction. Therefore, as one might expect, the oscillons should not appear. 

\begin{figure}
    \centering
   \hspace*{-0.5cm} \includegraphics[width=0.9\linewidth]{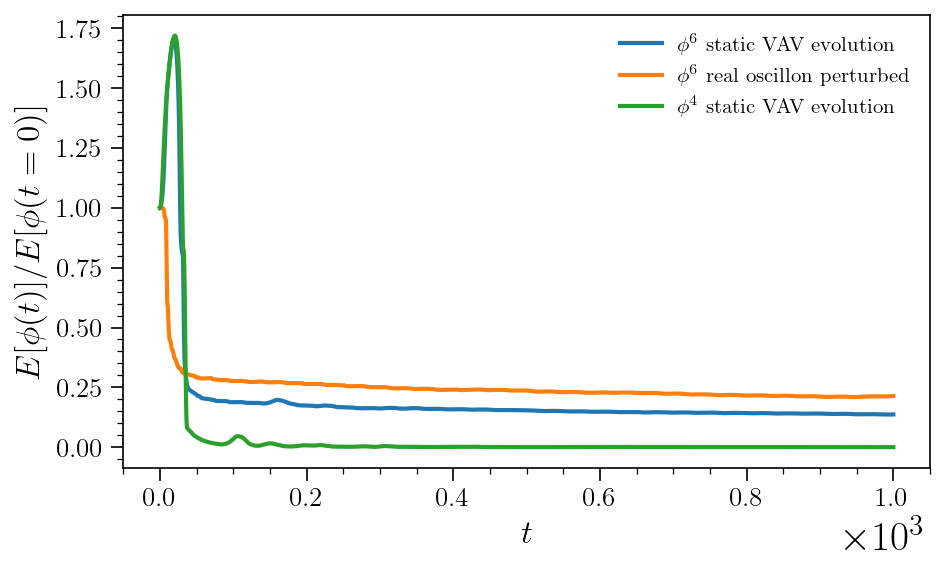}
    \caption{Evolution of the normalized energy enclosed in the disc with radius $R=d/2$ for the VAV collision with $v_{\rm in}=0$ in the $\phi^6$ (blue) and $\phi^4$ (green) models and for the perturbed Gaussian initial data (orange).}
    \label{fig:energy}
\end{figure}
\begin{figure}
    \centering
   \hspace*{-0.5cm} \includegraphics[width=1.0\linewidth]{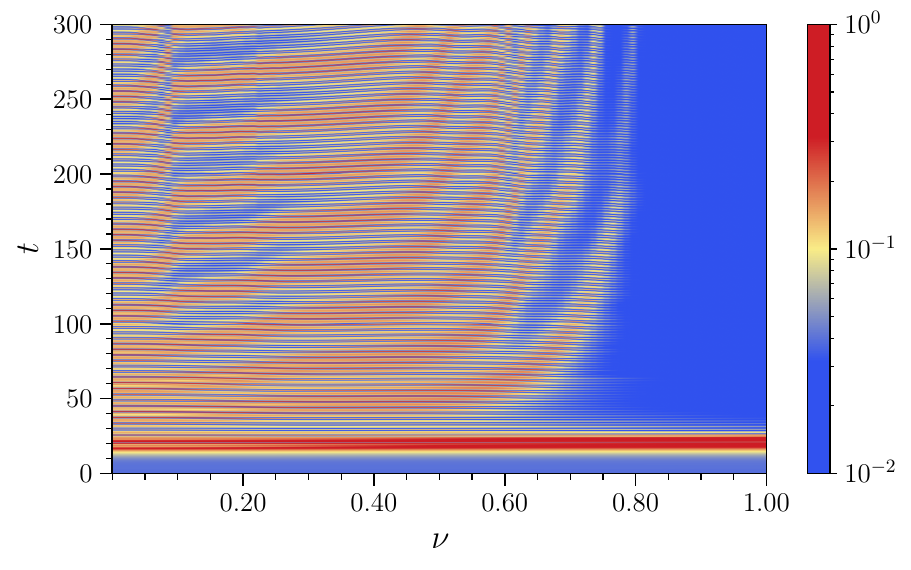}
    \caption{ The energy density at the origin for VAV collisions
as a function of time for various values of the parameter $\nu$ in potential (\ref{eq:V_6}). Here, $v_{in}=0$.}
    \label{fig:CL}
\end{figure}

The resulting oscillon is a strong attractor in dynamics of the complex $\phi^6$ theory rather than a solution generated by fine tuned initial data. It is formed during extremely violent processes where more than 80$\%$ of energy is radiated out, see Fig. \ref{fig:energy}, the blue curve. 

We found that the appearance of the large amplitude oscillon in the broken vacuum and, consequently, the resonant structure in VAV collisions crucially depends on the properties of the potential far away from this point. It seems that the existence of a sufficiently deep (false or true) unbroken vacuum at $\phi=0$ is a necessary condition. This happens for $\nu<1$, as can be seen in Fig. \ref{fig:CL}, where the oscillon formation occurs until $\nu \approx 0.75$. For larger values of $\nu$, we do not find any oscillons. 


\section{Broken vacuum oscillon in the complex $\phi^6$ model}

We will show that the oscillon formed in the VAV collisions is an embedded large amplitude real $\phi^6$ oscillon, that can be easily found in the real truncation of the $\phi^6$ model starting from a generic Gaussian initial data
\begin{equation}
    \phi(t=0,r) = \sqrt{2} \left(1 - A\,e^{-r^2/\sigma^2} \right). \label{gauss}
\end{equation}
Here, $A$ and $\sigma$ denote the amplitude and width of the initial pulse, respectively.

An approximate expression for oscillons in the real $\phi^6$ model can be derived using the small-amplitude expansion \cite{Fodor:2008es}, see the Appendix \ref{real-oscillon}. This is an unmodulated oscillon. In our numerics, we find excited oscillons that possess characteristic amplitude modulations.

\begin{figure}
    \centering
   \hspace*{-0.0cm} \includegraphics[width=0.9\linewidth]{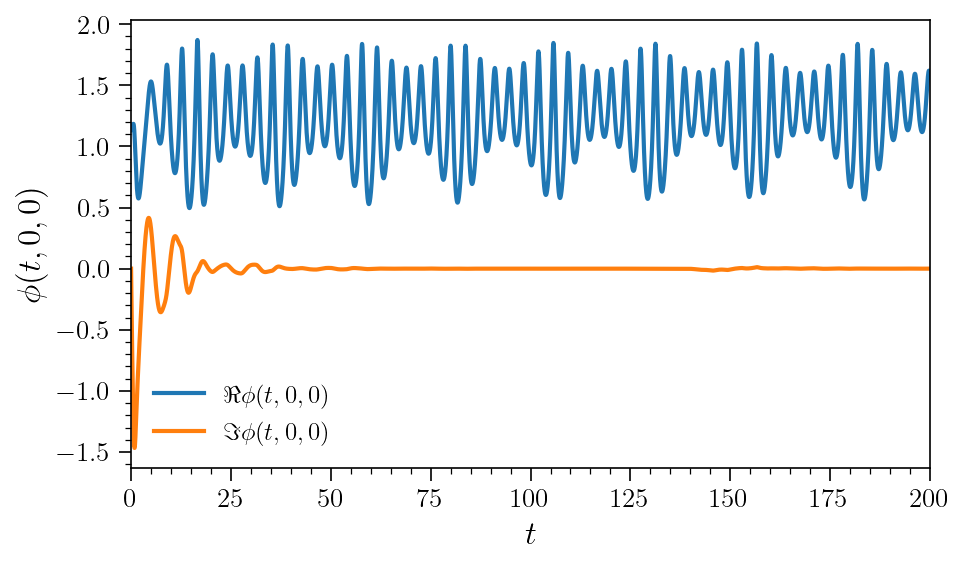}
    \caption{Embedded real $\phi^6$ (generated from the Gaussian data with $A=0.2$ and $\sigma=2.9$) with a perturbation along the imaginary component. The real and imaginary component of $\phi$ at the origin. }
    \label{fig:embedded}
\end{figure}

In Fig. \ref{fig:embedded} we show the evolution of the field at the origin generated from the Gaussian initial data (\ref{gauss}) perturbed along the imaginary direction. This is achieved by assuming a non-zero time derivative of the field
\begin{equation}
    \partial_t \phi(t=0,r)=3\, i\,e^{-r^2/\sigma^2} \left( \frac{2\sqrt{2}\,r}{\sigma^2} - \sqrt{\pi} \frac{2\sqrt{2} - A}{\sigma (2-A)} \right).
\end{equation}
This form of the perturbation yields a vanishing $U(1)$ charge and therefore corresponds to the initial conditions for the head-on VAV collisions. The field settles very quickly on the embedded real oscillon with the imaginary component radiated out. The real component oscillates very much as in the VAV case. 

\begin{figure}
    \centering
   \hspace*{-0.0cm} \includegraphics[width=0.9\linewidth]{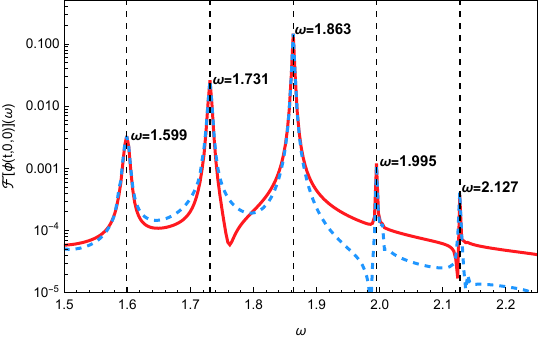}
    \caption{Power spectrum of the field at the origin: oscillon found in the VAV collision with $v_{\rm in}=0$ and $2d=20$ (red line); The embedded oscillon found from the perturbed Gaussian initial data with $A=0.2$ and $\sigma=2.9$ (blue dashed line).}
    \label{fig:power}
\end{figure}

The similarity between the VAV oscillon and the embedded modulated $\phi^6$ oscillon is particularly well visible if we compare the power spectra, see Fig. \ref{fig:power}. Within our numerical accuracy, the fundamental frequency and the frequency of the modulations are in both cases the same. This underlines the fact that this oscillon behaves as a very strong attractor in the evolution of the $\phi^6$ model and field settles down on it for very different initial data. Furthermore, as is typical for oscillons, the spectrum is $\mathbb{Z}_2$ symmetric, i.e., invariant under $\omega \to -\omega$.

In contrast, the real $\phi^4$ oscillon \cite{Salmi:2012ta} decays rapidly if it is embedded in the complex version of the theory and perturbed in an arbitrary way. The decay occurs mainly via radiation in terms of massless Goldstone waves. 

An oscillon also appears in the annihilation of local vortices in the Abelian-Higgs model deep in the type-I regime \cite{Gleiser:2007te}. Note that after the gauging, the model has a non-zero mass threshold both in the matter and in the gauge sector and therefore its existence is expected. 
\section{Conclusions}
In the present work, we have found that the addition of a higher order non-linear self-interaction term leads to an amazingly complicated pattern in the annihilation process of global vortices, as it was proved for a one-parametric family of the $\phi^6$ theories. This effect is a consequence of the surprising existence of an oscillon in the broken vacuum, where, due to the flat direction and lack of the mass threshold, no oscillons were expected. 

Even more surprisingly, the existence of the broken vacuum oscillon is strongly related to the behaviour of the potential {\it far away from the broken vacuum} around which this non-perturbative excitation oscillates. Precisely speaking, a sufficiently deep isolated local minimum, i.e., {\it a false vacuum}, (e.g., at $\phi=0$) seems to be a necessary condition. This structure of the vacua is a common feature of the SM extensions \cite{Espinosa:1993bs, Profumo:2007wc, Kurup:2017dzf, Lewicki:2021pgr}.


The existence or not of the long-lived oscillons in global vortex models has a huge impact on the long time evolution of a multi-VAV ensemble, which is expected to appear during a phase transition. It is clear that in the $\phi^6$ model such a system will be dominated by  oscillons, whereas in the $\phi^4$ theory all the localized structures will annihilate into radiation. This may have an important consequence for the evolution of axionic strings and the medium- and even large-time abundance of dark matter axions. Note that probably the oscillons can also arise from the decay of the global string loops.


The fate of the broken vacuum oscillons after quantization is an open problem. However, 
despite previous work \cite{Hertzberg:2010yz}, recently it has been shown that quantum oscillons do exist \cite{Evslin:2025hjt}. Hence,  they can  potentially exist in higher order operator extension of the usual Mexican hat potential.

Looking from a wider perspective, our research shows that non-perturbative properties of the theory (potential) far away from the broken vacuum where {\it we live} can significantly modify the particle content and dynamics in {\it our Universe}. Therefore, it would be very interesting to understand how higher order interaction terms modify dynamics of other solitons based on the broken vacuum, e.g., local vortices or monopoles. 
\section*{Acknowledgement}
The authors acknowledge support from the Spanish Ministerio de Ciencia e Innovacion (MCIN) with funding from the grant PID2023-148409NB-I00 MTM.  D.M.C. acknowledges financial support from the European Social Fund, the Operational Programme of Junta de Castilla y Le\'on, the regional Ministry of Education and the  Laboratory of Disruptive and Interdisciplinary Sciences
(CLU-2023-1-05).

The authors thank M. Bachmaier and T. Romanczukiewicz for discussion.

\vspace*{0.7cm}

\begin{center}
    \textbf{\large SUPPLEMENTAL MATERIAL}
\end{center}

\appendix
\section{Numerical setup} \label{numerical-setup}

For the numerical integration, we used two different methods, a finite-difference scheme and a fourth-order Yoshida's symplectic integrator \cite{Yoshida:1990}, discretizing the field equation in a box $\Omega=[-L,L]\times[-L,L]$. Additionally, we used fourth-order Mur boundary conditions as well as a damping term to ensure that the radiation emitted during the vortex/antivortex scattering does not propagate back to the centre of the simulation box. Hence, the field equation becomes
\begin{equation}\label{fieldEqDamping}
    \partial_{t,t}\phi+\gamma(x,y)\,\partial_t \phi =\nabla^2\phi -\frac{\partial V}{\partial\overline{\phi}}.
\end{equation}
Furthermore, in order to check numerical accuracy, we tried two different dampings,
{\scriptsize
\begin{equation}
    \gamma(x,y)=\left\lbrace\begin{matrix}
        \left(2\frac{{\mathrm max}(|x|,|y|)-l_{damp}}{L-l_{damp}}\right)^4, &  {\mathrm{if} } \,\,\mathrm{max}(|x|,|y|)>l_{damp},\\
        0, & {\mathrm{if} } \,\,\mathrm{max}(|x|,|y|)<l_{damp},
    \end{matrix} \right.
\end{equation}
}and
{\scriptsize
\begin{equation}
    \gamma(x,y)=\left\lbrace\begin{matrix}
       K(x,y), &  {\mathrm{if} } \,\,\mathrm{max}(|x|,|y|)>l_{damp},\\
        0, & {\mathrm{if} } \,\,\mathrm{max}(|x|,|y|)<l_{damp},
    \end{matrix} \right.
\end{equation}
}being $K(x,y)\equiv 1 - (1 - e^{-\alpha(x - L)^2})(1 - e^{-\alpha(y - L)^2})$, see \cite{Krusch:2024vuy}. In both cases, the damping term is non-zero only near the boundaries of the simulation box. 

For the finite-differences case, the field equation was discretized 
using a fourth-order finite-difference scheme in space and a 
second-order finite-difference scheme in time so that the field equation can be written as
{\small
\begin{eqnarray*}
 & &   \frac{\phi^{n+1}_{i,j}+\phi^{n-1}_{i,j}-2\phi^{n}_{i,j}}{dt^2}+\gamma(x_i,y_i) \frac{\phi^{n+1}_{i,j}-\phi^{n-1}_{i,j}}{2\, dt}=\\
   & & \frac{-\frac{\phi_{i-2,j}^{n}}{12}+\frac{4\phi_{i-1,j}^{n}}{3} - \frac{5\phi_{i,j}^{n}}{2}+\frac{4\phi_{i+1,j}^{n}}{3}-\frac{\phi_{i+2,j}^{n}}{12}}{dx^2} + \\
   & & \frac{-\frac{\phi_{i,j-2}^{n}}{12}+\frac{4\phi_{i,j-1}^{n}}{3} - \frac{5\phi_{i,j}^{n}}{2}+\frac{4\phi_{i,j+1}^{n}}{3}-\frac{\phi_{i,j+1}^{n}}{12}}{dy^2} -2\frac{\partial V_6}{\partial\overline{\phi^{n}_{i,j}}},
\end{eqnarray*}
}where the indices $i,j$ label the spatial grid points and $n$ labels the time step. 
At each time step we compute the value of $\phi_{i,j}^{n+1}$ at every grid point.

\begin{figure}
    \centering
   \hspace*{-0.5cm} \includegraphics[width=1.0\linewidth]{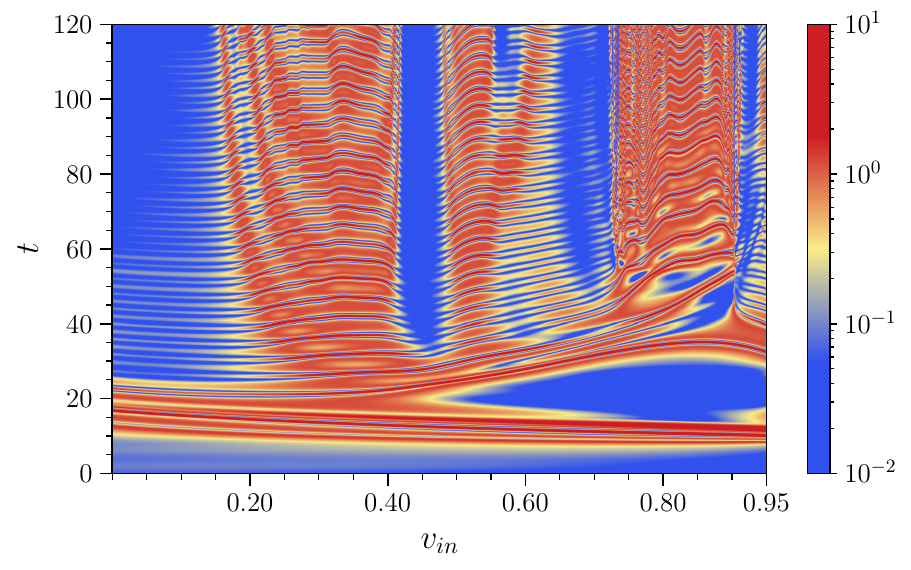}

\centering
      \hspace*{-0.5cm} \includegraphics[width=1.0\linewidth]{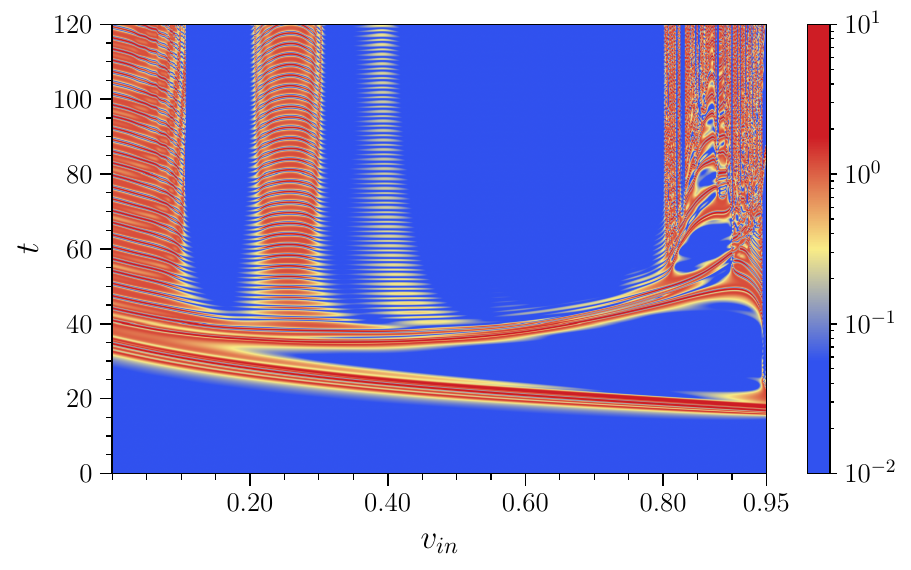}
    \caption{Energy density at the origin as a function of time for various values of the initial velocity $v_{\rm in}$ for initial separation $2d=15$ (upper) and $2d=40$ (lower).}
    \label{fig:vav-more-exmaples}
\end{figure}

The results presented in this article have been obtained using a spatial grid spacing $dx=dy=0.1$, and time 
step $dt=0.01$. Additionally, we chose $l_{\text{damp}}$ such that the damping only affects the $10\%$ of the boundary. Nevertheless, we also checked for different steps and spacings with no significant changes in the results.

As initial conditions for a colliding VAV pair separated by a distance $d$, we use the Abrikosov ansatz;
{\scriptsize
\begin{eqnarray}
   \hspace{-0.2cm} \phi(x,y, 0)\hspace{-0.2cm}&=&\hspace{-0.2cm}\frac{1}{\sqrt{2}}\phi_+(\gamma(x + d),y)\,\phi_-(\gamma(x -d),y), \label{initial-data1}\\
     \hspace{-0.2cm}   \phi(x,y, dt)\hspace{-0.2cm}&=&\hspace{-0.2cm}\frac{1}{\sqrt{2}}\phi_+(\gamma(x-v \, dt + d),y)\,\phi_-(\gamma(x+v \, dt -d),y). \label{initial-data2}
\end{eqnarray}
}
being $v$ the initial velocity.

Within the highly relativistic regime, $v_{\rm in}>0.85$, the lattice resolution could be insufficient. Nevertheless, we expect that qualitatively the results will not change. 
\section{Oscillon from the VAV pair} \label{VAV-d}

The VAV configuration (\ref{initial-data1}), (\ref{initial-data2}) can be treated as a family of initial profiles parametrized by the parameter $d$. Depending on its value, we have a more or less separated VAV pair, that is, an initial pulse with larger or smaller energy, depending on whether $d$ grows or decreases. 

In Fig. \ref{fig:vav-more-exmaples}, we show the resulting time evolution of the energy density at the origin for different values of the initial velocity for small $2d=15$ and large separation $2d=40$. In both cases, regimes with the creation of oscillons (red regions) are clearly visible. Again, they are chaotically separated by bounce windows which end up in an oscillon or in annihilation to the vacuum. 

Importantly, for a large set of the values of the initial parameters ($d$ or $v_{\rm in}$), we basically  find the same modulated oscillon with the same $\omega_0$ and $\omega_{\rm mod}$, which shows that this solution is an extraordinarily strong attractor in the dynamics of the complex $\phi^6$ model.

Another observation is that once the energy stored in the initial configuration decreases below a critical value we do not find the oscillon. On the one hand, such initial data have too little energy to find the large amplitude modulated oscillon. On the other hand, it shows that small oscillons do not exist in the complex $\phi^6$ extension. 

\section{Bounce windows} \label{VAV-bounce}
\begin{figure}
    \centering
    \includegraphics[width=0.49\linewidth]{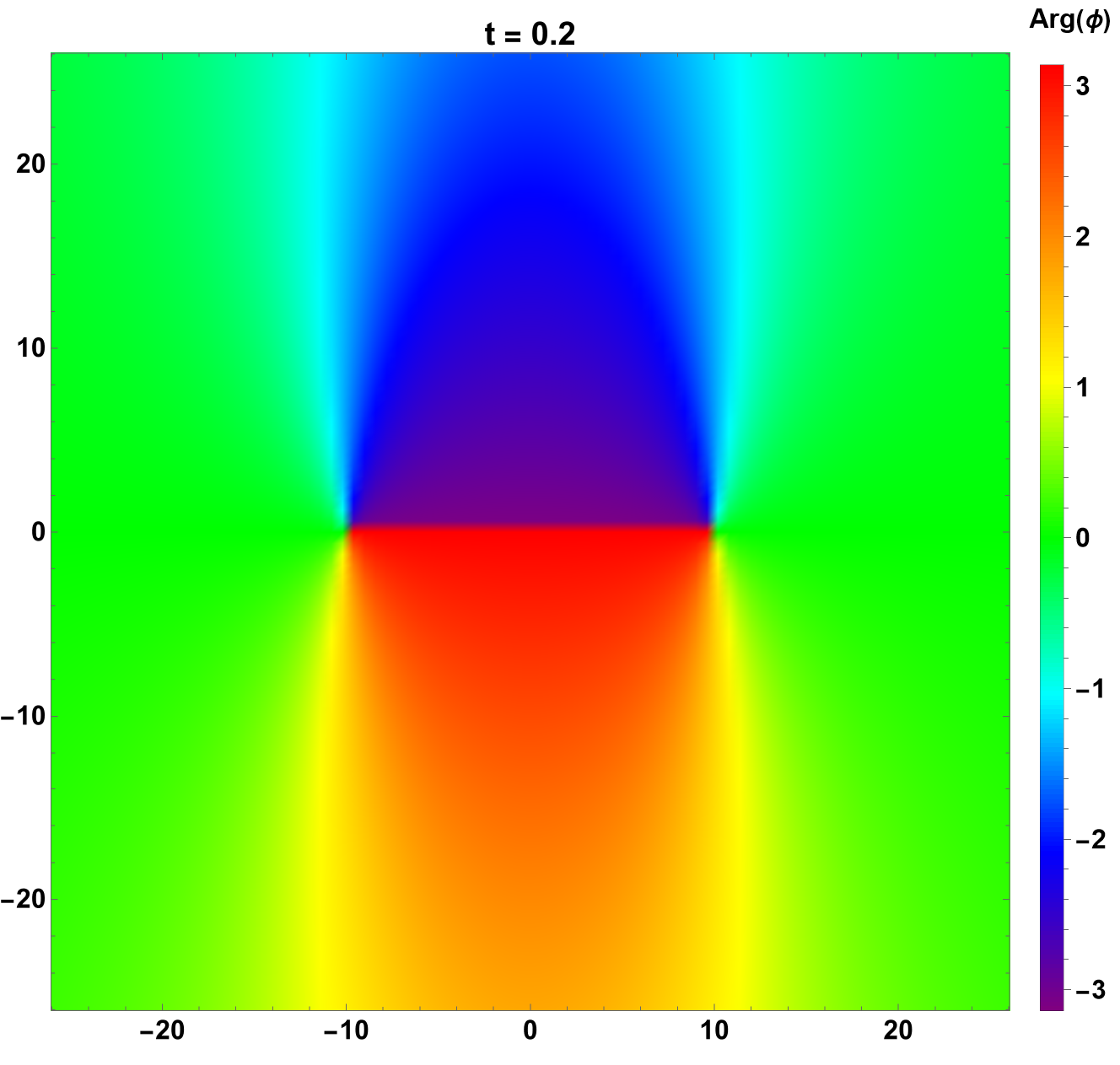}
    \includegraphics[width=0.49\linewidth]{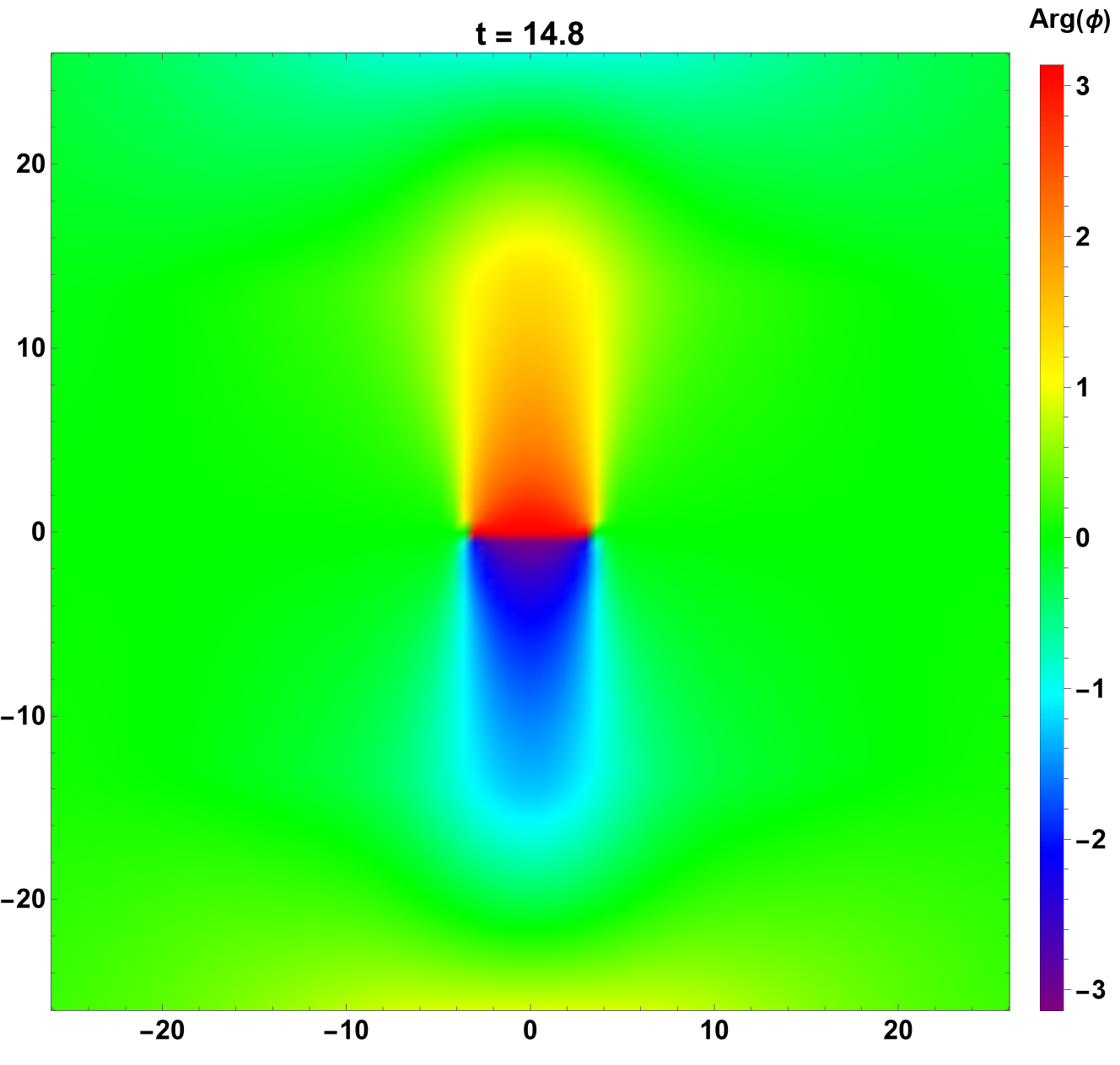}
    \includegraphics[width=0.49\linewidth]{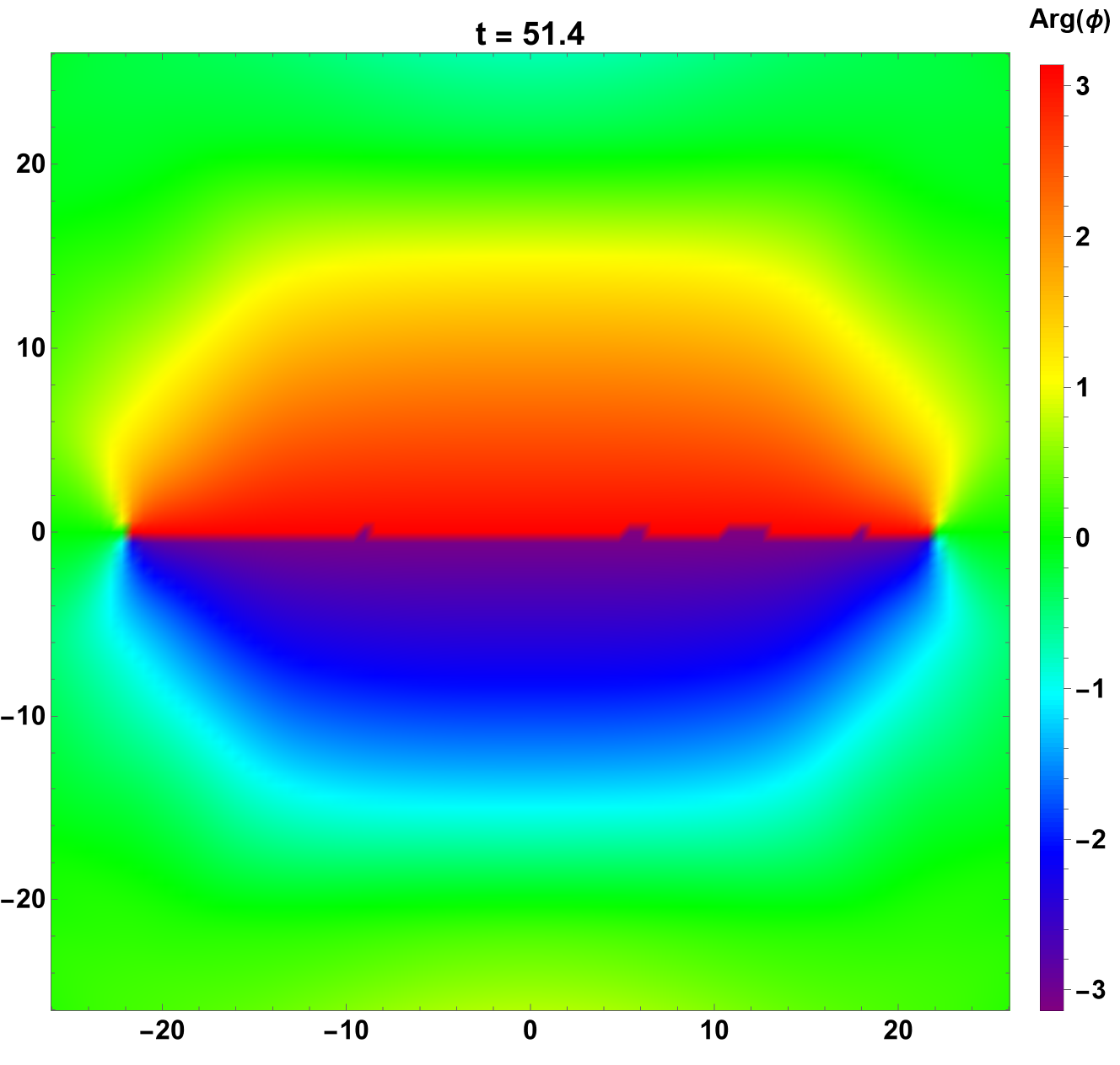}
    \includegraphics[width=0.49\linewidth]{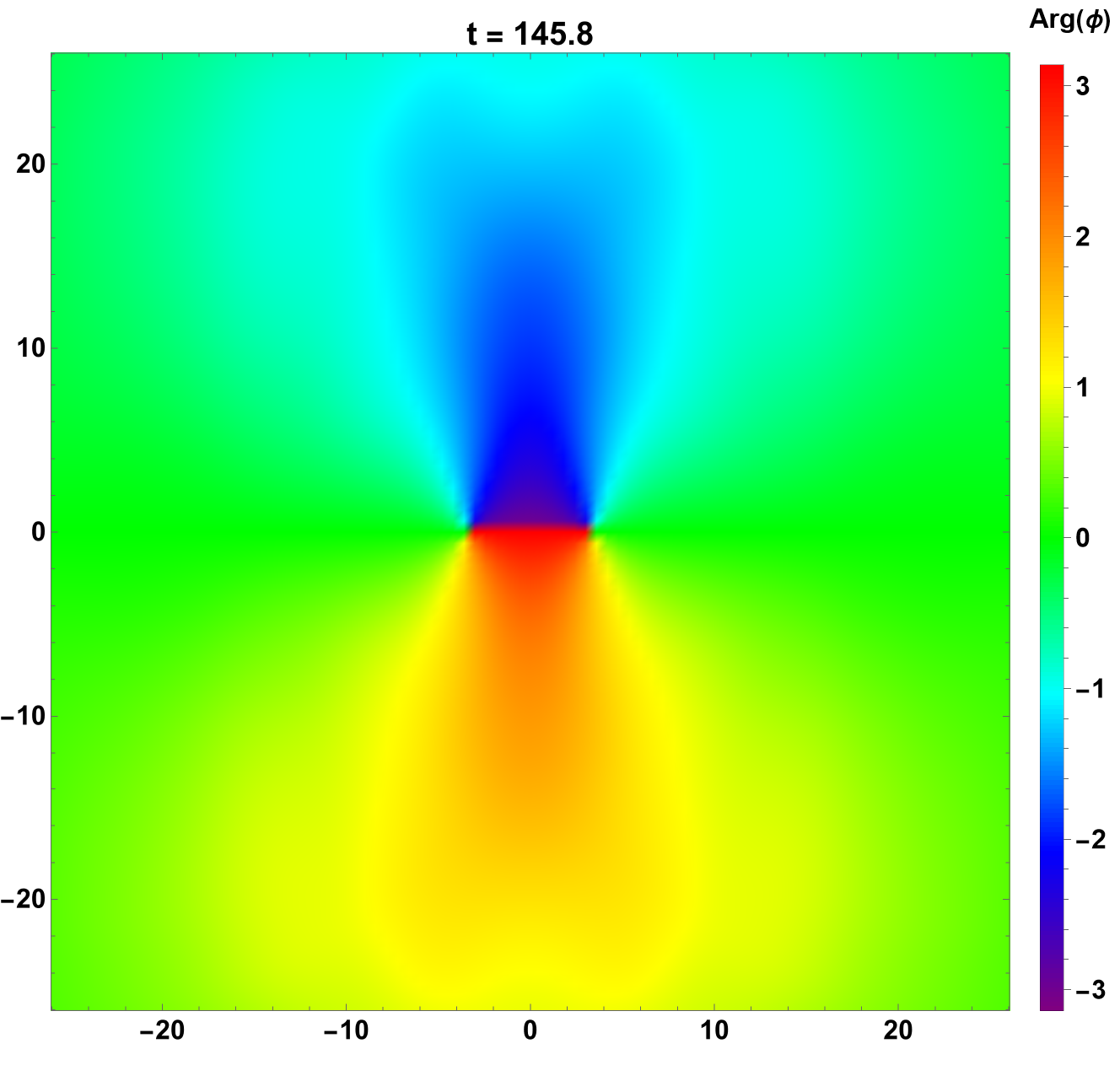}
    \caption{Phase of the field of a 3 bounce solution in the VAV collision in the $\phi^4$ model before the first collision ($t=0.2$), after the first collision ($t=14.8, 51.4$) and after the second collision ($t=145.8$). The third collision eads up in a direct annihilation. Here, $2d=20$ and $v_{\rm in}=0.95$. }
    \label{fig:vav-phase}
\end{figure}

Here, we show that in bounce windows, where the vortices are temporary recreated after the collision, they interchange their positions. Thus, we observe passing through scatterings rather than backscattering. To illustrate this, we plot the phase of the complex field in the plane, Fig. \ref{fig:vav-phase}. 

It is clearly visible that initially we have antivortex on the right hand side and vortex on the left hand side. After the first collision, the vortices go through each other. The same happens after the second collision. The third collision ends up in a direct annihilation. 

The passing through scenario is a feature of both $\phi^4$ and $\phi^6$ models. This is in agreement with the VAV scatterings in the Abelian Higgs model with the large value of the coupling constant $\lambda$. Here, for sufficiently high velocities, the vortices do not annihilate but pass through each other \cite{Bachmaier:2025igf}. 
\section{Modes of the global $\phi^6$ vortex} \label{modes}
The structure of the modes of the unit charge $\phi^6$ vortex is qualitatively very similar to the case of the $\phi^4$ theory \cite{Blanco-Pillado:2021jad}. Let us consider the simplest radial perturbation of the unit charge $\phi^6$ vortex 
\begin{equation}
    \phi(r,\theta)=f(r)e^{i\theta}+\eta(r) e^{i\theta} \cos \omega t,
\end{equation}
where $\eta$ is a small perturbation. After inserting it into the equation of motion and keeping only the linear terms in $\eta$ we arrive at the following eigenvalue problem 
\begin{equation}
- \frac{1}{r} \frac{d}{dr} \Big( r \frac{d \eta}{dr} \Big)
+ \left(  \frac{1}{r^2}+\frac{15}{4} f^4 - 6 f^2 + 1 \right) \eta
= \omega^2 \eta,
\end{equation}
where the $1/r^2$ core in the Schr\"{o}dinger potential comes from the non-zero winding number. $\omega$ is the frequency of the eigen-mode. Introducing a new function $\psi$, such that $\psi = \sqrt{r} \eta$, we can rewrite the equation as a one-dimensional Schr\"{o}dinger problem
\begin{equation}
- \frac{d^2 \psi}{dr^2} +V_{eff} \psi = \omega^2 \psi,
\end{equation}
where the effective potential 
\begin{equation}
    V_{eff} (f)= \frac{3}{4r^2} + \frac{15}{4}f^4 -6f^2+1.
\end{equation}
The asymptotic behavior of the vortex profile for $r\to \infty$ is
\begin{equation}
    f(r)=\sqrt{2} - \frac{\sqrt{2}}{4r^2} +o(r^{-2})
\end{equation}
Hence, the potential in the spectral problem has the following asymptotic form
\begin{equation}
   V_{eff}(r)=  4- \frac{33}{4r^2}  +o(r^{-2}). 
\end{equation}
The appearance of a long-range $r^{-2}$ tail in the spectral problem potential results in infinitely many bound states accumulating at the mass threshold \cite{Blanco-Pillado:2021jad}.

We found the five lowest modes. They have frequencies: $\omega_1=0.535,\, \omega_2=1.696,\, \omega_3=1.978,\, \omega_4=1.998, \,\omega_5=1.999$, which do not correspond to the fundamental frequency of the oscillon.  

Importantly, during collisions, we do not observe significant excitation of these modes. In particular, between the first and second bounce, in two-bounce collisions, we did not measure frequencies, which would correspond to the internal oscillations of the single vortex. Thus, the resonant structure is certainly not triggered by these modes but, as claimed in the paper, by the oscillon. 

We should remember that, contrary to the modes of local vortices \cite{Goodband:1995rt,Alonso-Izquierdo:2015tta, Alonso-Izquierdo:2024bzy, Gavrea:2026adu}, the modes of global vortices are Feshbach resonances \cite{Feshbach}. These are modes in a multi-channel spectral problem, where each channel has a different mass threshold. Then, the mode can be a bound mode in one channel (with the frequency $\omega$ below the mass threshold) whereas, in another channel, it can have an unbounded scattering-like component ($\omega$ is above the mass threshold of this channel). Here, the second channel, is the phase of the complex field, where there is no mass threshold at all. 

The half-bound excitation is known to be quite stable, with a slow decay into the radiative component. Hence, such a resonance may have a rather significant impact on the dynamics of solitons, see, e.g., \cite{GarciaMartin-Caro:2025zkc} and \cite{Bachmaier:2025igf}. Interestingly, they are the modes of the BPS t'Hooft-Polyakov monopole \cite{Forgacs:2003yh, Russell:2010xx}. 

\section{Small oscillons in the real $\phi^6$ model} \label{real-oscillon}
A small amplitude symmetric oscillon in the $\phi=0$ vacuum in the $\phi^6$ model has been found in \cite{Fodor:2009kf}. Repeating this procedure \cite{Fodor:2008es}, the first order approximation to the asymmetric oscillon in the $\phi=\sqrt{2}$ vacuum reads
\begin{equation}
    \Phi(t,r)=\sqrt{2}+\frac{\epsilon}{2\sqrt{\lambda_F}} S\left(2r\right) \cos (\sqrt{4-\epsilon^2}t), \label{Fodor}
\end{equation}
where the profile function $S(r)$ obeys
\begin{equation}
    \frac{d^2S(r)}{dr^2} +\frac{1}{r} \frac{dS(r)}{dr} -S+S^3=0 \label{master}
\end{equation}
and $\lambda_F=6$. $\epsilon$ is the expansion parameter, which is assumed to be small. The numerical solution of the master equation (\ref{master}) is presented in \cite{Fodor:2008es}. In particular $S(0)=2.206$. 

If we apply this approximation to the frequency of the Feshbach oscillon, we find $\epsilon=\sqrt{4-\omega_0^2}\approx 0.786$. This allows us to compute the amplitude of the oscillations $\Delta\Phi \approx \pm 0.35$. This crudely agrees with the amplitude at the minimum of the modulations. At maximum, the observed amplitude is much larger. Of course, the small amplitude approximation does not capture the amplitude modulations at all. 


\bibliography{ref}

\end{document}